# Artificial intelligence for simplified patient-centered dosimetry in radiopharmaceutical therapies

Alejandro Lopez-Montes [a]*, Fereshteh Yousefirizi [b], Yizhou Chen [a], Yazdan Salimi [c], Robert Seifert [a], Ali Afshar-Oromieh [a], Carlos Uribe [b,d], Axel Rominger [a], Habib Zaidi [c], Arman Rahmim [b,d], Kuangyu Shi [a]

*a Department of Nuclear Medicine, Inselspital, Bern University Hospital, University of Bern, Switzerland*
*b Department of Integrative Oncology, BC Cancer Research, Vancouver, BC, Canada*
*c Division of Nuclear Medicine and Molecular Imaging, Geneva University Hospital, Geneva, Switzerland*
*d Department of Molecular Imaging and Therapy, University of British Columbia, Vancouver, BC, Canada*

* Corresponding author: Alejandro Lopez-Montes; alejandro.lopez@unibe.ch

**KEY POINTS:**

- The rapid evolution of radiopharmaceutical therapy (RPT) highlights the growing need for personalized and patient-centered dosimetry.
- Artificial Intelligence (AI) offers solutions to the key limitations in current dosimetry calculations.
- The main advances on AI for simplified dosimetry toward patient-friendly RPT are reviewed.
- Future directions on the role of AI in RPT dosimetry are discussed.

## INTRODUCTION

Dosimetry is crucial for understanding the response of both organs at risk (OARs) and tumor lesions in radiopharmaceutical therapy (RPT) [1,2]. Accurate dosimetry can facilitate enhanced treatment design for cancer patients undergoing RPT [3,4]. Different patients present unique anatomies and response to radiation treatments [5,6]. However, the injected dose for RPT is often fixed according to pre-established fixed protocols which do not include individual considerations [6,7].

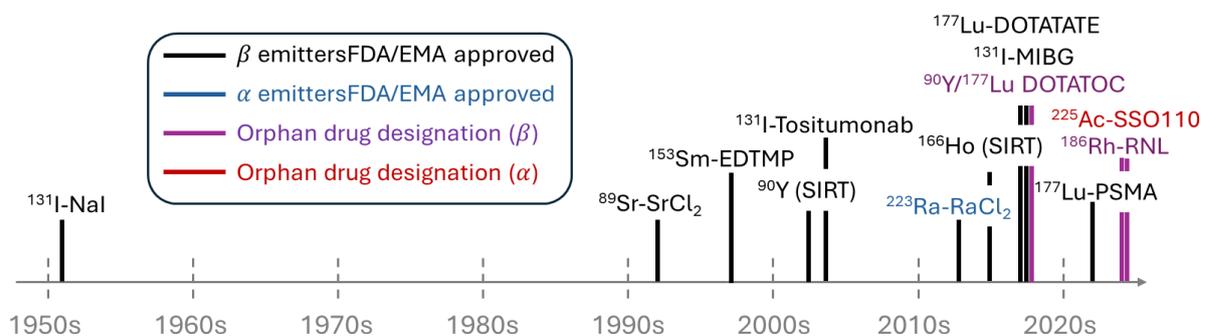

**Fig.1.** Evolution of radiopharmaceuticals with FDA or EMA approval for commercialization for therapeutical applications as of June 2025.

Over the past decade, RPTs have undergone significant growth [8,9]. Several $\beta^-$ (electron) emitters have been used to radiolabeled drugs such as $^{177}$Lu-PSMA-617 for the treatment of metastatic castrate-resistant prostate cancer (mCRPC) [10,11] and $^{177}$Lu-DOTATATE which is approved for neuroendocrine tumors [12,13]. Several additional drugs are already approved (see Fig.1) while others are currently under review by regulatory agencies [14–19]. Some others are under a fast-track approval by the FDA. Among them, $\alpha$-emitting radiopharmaceuticals are of particular interest: $^{225}$Ac-PSMA for prostate cancer [20,21]; $^{225}$Ac-ABD-147 for lung and neuroendocrine tumors [22]; or $^{212}$Pb-VMT01 for metastatic melanoma [23,24]. One of the benefits of $\alpha$ emitters is their increased stopping power and energy deposition compared to $\beta^-$, which make them ideal candidates for targeting tumor cells while sparing healthy tissues. Other compounds in clinical trials, such as $^{161}$Tb-PSMA ($\beta$ decay with additional Auger electron emission), present very promising applications with potentially enhanced therapeutic properties for mCRPC [25,26].

This growth and the current lack of truly personalized treatments have motivated the need for more reliable and patient-centered approaches [27,28]. However, dosimetry in internal radiotherapy remains challenging due to the dynamic bio-distribution of the radiopharmaceuticals [5,29] and complex models and time-consuming protocols are often required for accurate absorbed dose estimation [29]. Consequently, simplified strategies are essential to enable the broader clinical implementation of personalized dosimetry in RPT [27,28]. Conventional strategies have shown insufficient to address these limitations. Artificial Intelligence (AI) strategies, including Machine Learning (ML) and Deep Learning (DL), present promising solutions (Fig.2) with special interest on patient-friendly approaches designed to achieve truly personalized dosimetry [30]. In this review, we summarize the main advances in AI towards a simplified patient-centered dosimetry in RPT.

**CLINICAL RELEVANCE OF PATIENT-FRIENDLY DOSIMETRY**

Patient-specific dosimetry enables treatment based on the patient size, tumor extension and/or risk profile [1,8,31]. The optimization of the tumor dose often translates into adjusting the administered activity and determining the number of cycles to maximize lesion control while minimizing exposure to healthy organs [27]. This principle also applied to benign thyroid disease, where the administered activity of $^{131}$I depends on the functional state of the thyroid, the size and uptake of the autonomous nodules, and disease-specific factors (e.g. ablative concept for Graves' disease [32]). In contrast, for thyroid cancer, the activity of $^{131}$I administered for remnant ablation and/or adjuvant therapy is commonly based on fixed activity schemes that vary according to disease severity (e.g., presence of bone metastases vs. suspected local recurrence) [33]. To optimize the thyroid cancer tumor dose, some centers leverage [124]I PET to determine the optimal activity of [131]I that can be safely used for treatment [34].

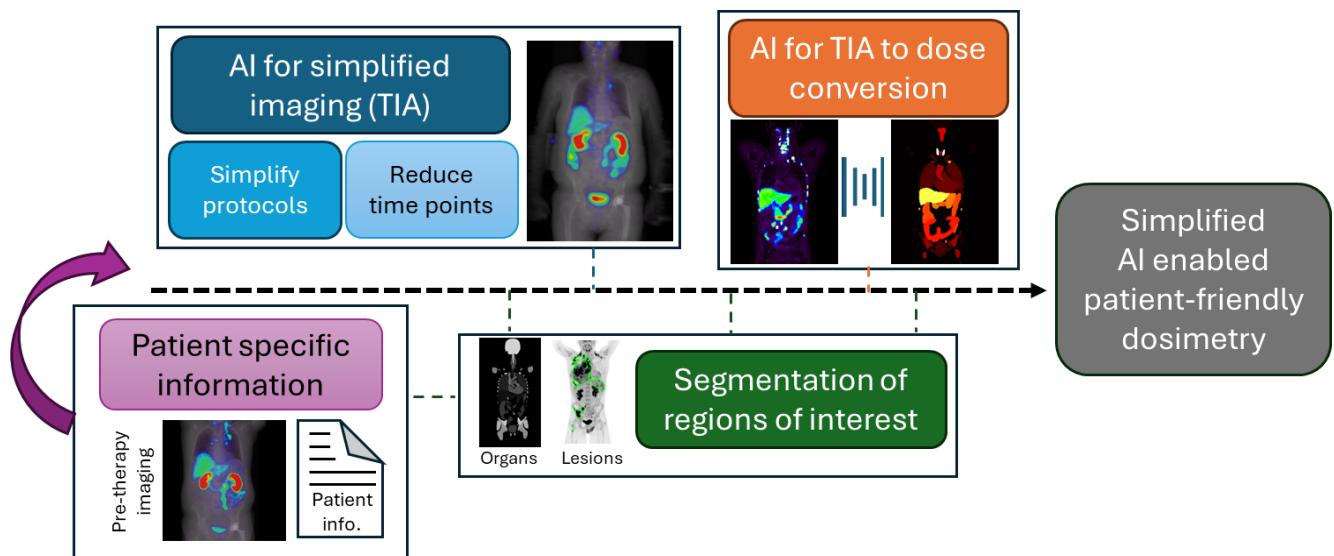

**Fig.2.** Patient-friendly dosimetry and the role of AI for simplified strategies.

For radioembolization of liver cancer, such as hepatocellular carcinoma (HCC), with selective internal radiation therapy (SIRT), tailoring the administered activity to the patient is essential. The DOSISPHERE-01 trial (using [90]Y glass microspheres) targeted a minimum tumor-absorbed dose of ≥ 205 Gy and demonstrated superior outcomes versus standard dosing [35]. Note that this threshold is product-specific and studies with resin microspheres report lower tumor-dose thresholds (~100–150 Gy)**,** depending on dosimetry method and clinical endpoint [36]. Following this, optimization of tumor dose is incorporated into the EANM guidelines and clinical practice for SIRT [19,37]. In contrast, radioligand therapies like peptide receptor radionuclide therapy (PRRT) for neuroendocrine tumors, or PSMA therapy for prostate cancer commonly involve fixed administered activities [38], with adjustments primarily reserved for patients with higher risk of hematological toxicity [39]. This is the case even though a dose escalation study has shown that higher activity of PSMA therapy is feasible [40]. However, the dosimetry (i.e. optimization of administered activity) for PSMA therapy is more complex compared with, for example, SIRT [31]. The reason for this is twofold: first, it is not clear which clinical implication a certain dose has for the tumor or organ at risks. The dose-response relationships have been primarily estimated based on external beam radiation, which is not directly transferable, for example, to the $\beta$-emitting [177]Lu, which is primarily used for PSMA or DOTATATE therapies [41]. In addition, both predictive and post-treatment dosimetry are technically complex and thus not routinely performed [42]. These limitations are self-reinforcing, as the lack of systematic dosimetric data hinders the establishment of reliable dose-response models in radioligand therapy. Therefore, simplified dosimetry approaches for radioligand therapies are needed to address the existing information gap.

**DOSIMETRY IN RPT**

The main protocols for dose calculation in RPT were established by the Committee of Medical Internal Radiation Dose (MIRD) of the Society of Nuclear Medicine and the International Commission on Radiological Protection (ICRP) [29,43]. The absorbed dose in a target region, $D(r_T)$, can be estimated in two steps:

1. Estimate the time-integrated activity (TIA) during treatment at a source region ($r_s$)

$$\tilde{A}(r_s) = \int_0^\infty A(r_s, t)dt \qquad (1)$$

2. Convert TIA to absorbed dose using S-values (dose-conversion factors) [29].

$$D(r_T) = \sum_{r_S} \tilde{A}(r_s) \cdot S(r_S \rightarrow r_T, t) \cdot dt \quad (2)$$

Where $\tilde{A}(r_s)$ is the TIA for every source region ($r_s$), and $S(r_S \rightarrow r_T, t)$ are conversion factors (S-values) from the TIA in the source regions to the dose deposited in the target regions ($r_T$). Note that source and target regions may refer to any region from individual voxels (voxel-wise dosimetry) to specific organs (organ-wise dosimetry). Both TIA and S-values calculations involve either complex computations or broad assumptions that limit their accuracy [8,43,44]. S-values depend on the type and energy of the particles emitted by the radiopharmaceuticals and in the tissue composition of the body [43,45,46]. Monte Carlo (MC) simulations are the gold standard to model them via either the use of pre-calculated values on anthropomorphic phantoms [47,48] (Fig.3.A) or with patient specific simulations (Fig.3.B) [44,49]. On one hand, tabulated S-values are often not realistic enough, ignoring the individual anatomy of the patient [8,44,49]. Also, S-values calculated at organ levels are assumed to be homogeneous organ-wise, limiting their applicability for voxel-wise dosimetry [50]. On the other hand, patient-specific simulations can provide accurate dose estimation for arbitrary target and source regions, but they are usually time consuming [51]. GPU parallelization provides considerable acceleration in MC simulations [51,52], although it still requires many simulated particles and is not as widely available. Intermediate approaches based on kernel convolutions (sometimes referred as voxel S-Values or VSV) from precalculated MC simulations are a common choice (Fig.3.C) [49] which despite their efficacy, they can present inaccuracies, especially in the edges of different tissues [49]. AI solutions (Fig.3.D) promise an alternative for detailed, patient-specific dosimetry without unrealistic simplifications and with lower computational burden than full MC approaches.

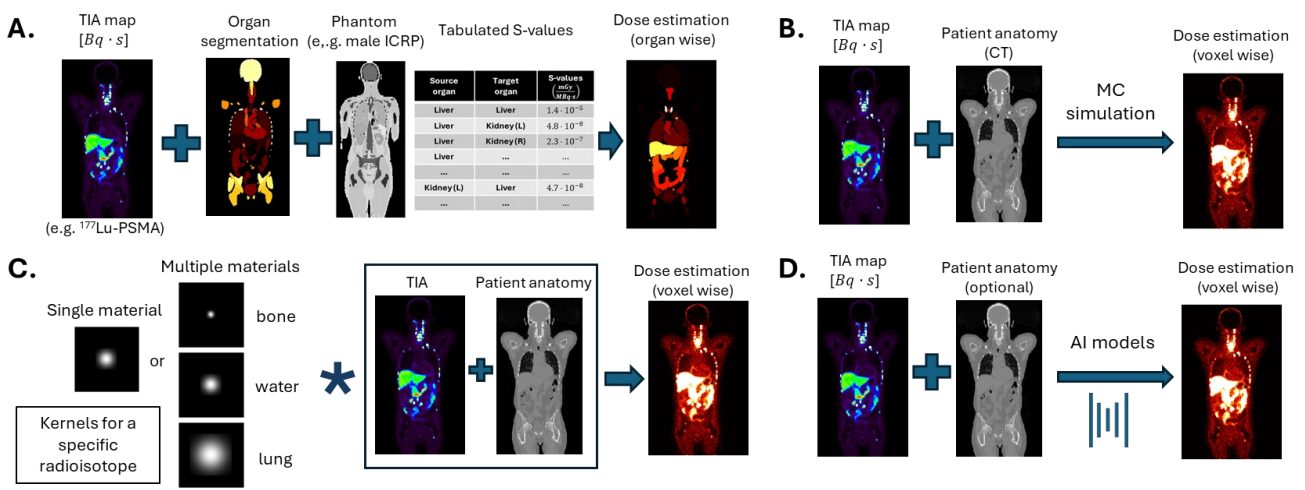

**Fig.3.** Example of dosimetry calculations from TIA maps using different strategies. **A.** Tabulated values from anatomical phantoms (ICRP in the example). **B.** MC simulation for a specific patient anatomy. **C.** Kernel based strategies based on pre-computed realistic simulations of the dose delivered by point sources of different radioisotopes in different media. **D.** AI solutions

Accurate estimation of the time-integrated activity (TIA, $\tilde{A}$) is essential regardless of the subsequent dose-conversion method. To obtain the TIA, serial quantitative imaging using Single Photon Emission Computed Tomography (SPECT) or, less ideally 2D-planar scintigraphy images [53,54]. Both modalities detect photons emitted directly by radionuclides used in RPT. However, SPECT is preferred due to its inherent 3D nature [54]. Sometimes, the use of theranostic pairs can be beneficial. In those cases, one isotope is used for treatment and another one for imaging, for example an equivalent PET tracer (Fig 4) [55,56]. The computation of the TIA often needs multiple time-point images to fit and calculate the area under the time-activity curve (TAC) in every source region [29,54]. The TAC can be modelled as an exponential decay with two half-lives: a physical one, $T_{1/2_{phys}}$, known and given by the radionuclide decay; and a biological half-life, $T_{1/2_{bio}}$, unknown and dependent on the pharmacokinetic distribution of the drug. The initial activity $A(r_s, 0)$ for every region is also unknown [29].

$$TIA = \tilde{A}(r_s) = \int_0^\infty A(r_s, t) \cdot dt;$$

$$A(r_s, t) = A(r_s, 0) \cdot e^{-ln\,2 \cdot t\left(\frac{1}{T_{1/2_{phys}}} + \frac{1}{T_{1/2_{bio}}}\right)} \quad (3)$$

Both parameters, $A(r_s, 0)$ and $T_{1/2_{bio}}$ can be estimated by curve fitting from different time-point observations $A(r_s, t)$. The number and moment in time at which the observations are taken are critical for accurate activity estimation [29,54,57]. The mono-exponential approach in (3) is sometimes replaced by multi-exponential models to consider more realistic variations of activity [58]. More detailed methodologies based on physiologically based pharmacokinetic (PBPK) modelling have also been proposed for the estimation of the TIA [59].

Thus, we can classify AI approaches for personalized dosimetry between solutions to simplify dose conversion from activity and others simplifying the imaging burden. Furthermore, segmentation and identification of lesions play an important role in every step of dosimetry calculations. Dosimetry estimation can also benefit from leveraging previous information about the patients, obtained from pre-therapy medical imaging or from patient reports. In the following sections, we review state-of-the-art AI-based solutions for each of these components of simplified, patient-centered dosimetry.

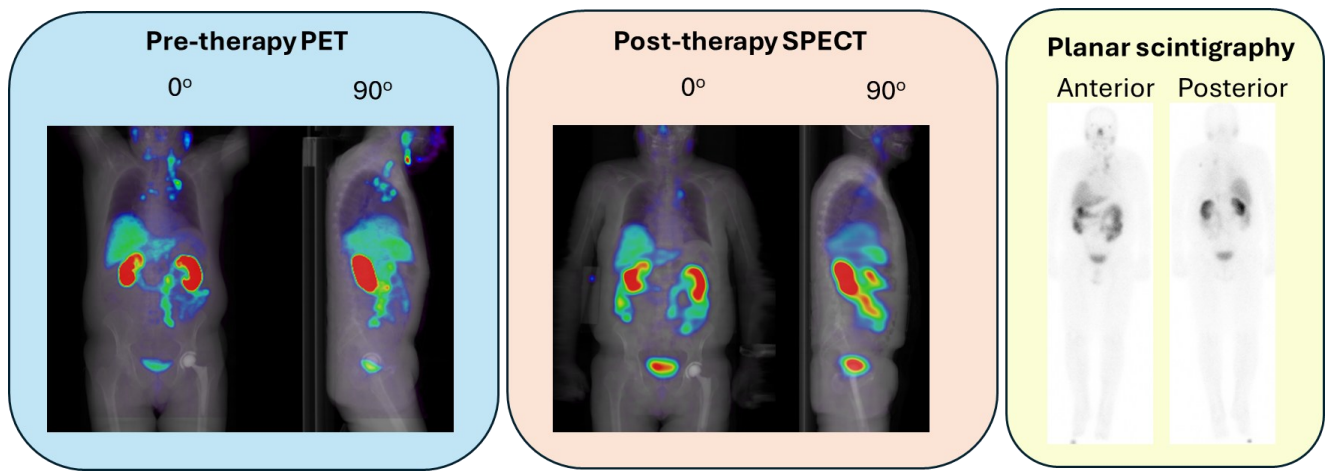

**Fig.4.** Example of different imaging modalities with relevance in dosimetry for the same patient. A. Pre-therapy [18]F-PSMA PET/CT. B. Post therapy [177]Lu-PSMA SPECT/CT obtained from 3 bed positions. A and B show projections at 0 and 90 degrees. C. Post therapy [177]Lu-PSMA planar scintigraphy (anterior and posterior) from a dual head gamma camera.

**SIMPLIFIED IMAGING PROTOCOLS FOR DOSIMETRY USING AI**

*Shortening imaging protocols*

SPECT can be a costly procedure due to long acquisition protocols; whole-body SPECT often requires multiple bed positions over ~40–60 min[60]. This imposes a burden on patients, with extended hospital visits at each time point . Therefore, a promise of AI for dosimetry is to reduce the burden of SPECT by limiting scanning time [61]-either by (i) reducing counts per projection (shorter dwell time) or (ii) reducing the number of projections via sparse-view SPECT (Fig. 5).

Both solutions present some challenges such as increased levels of noise in the reconstructed images or artifacts arising from insufficient sampling. Furthermore, SPECT imaging is challenged by worse resolution signal-to-noise ratio than other modalities like PET due, in part, to the need of collimators [56]. Several AI solutions have been proposed in recent years to tackle those limitations [61–63]. Wikberg et. al. and Ryden et. al. [60,64] proposed a reduction in the number of projections for [177]Lu SPECT from the standard 120 views to 30 [60] through the synthesis of 90 intermediate projections via Convolutional Neural Network (CNN). Using that approach, they reported comparable values between the full SPECT and the reduced number of views (SSIM values > 0.9) for at least the acquisitions during the first two days after injection [90]. Very similar approaches were suggested to recover full sinograms from sparse views in different applications of SPECT [65–67]. In [67], for instance, authors proposed a stationary SPECT system dedicated to cardiac imaging combined with a DL approach to increase sampling. Other approaches are based on the restoration of SPECT images after reducing acquisition time per projection [67–70]. For example, Lee et al. [69] proposed a strategy based on sinogram restoration from noisy data based on semi-supervised learning and leveraging inner-structure of the sinograms. Similar observations have been made by other works, as in Qi et al. [70], where the authors developed an image-to-image CNN to recover the quality of 20 min standard SPECT (120 views) from inputs of 3 mins SPECT reducing the time per frame. They achieved very comparable results between the fast SPECT protocol with DL enhancement and the standard protocol in [99m]Tc bone scans. While not directly proposed for dosimetry, other AI approaches to simplify SPECT for applications such as cardiac imaging [71,72] can have a direct translation for RPT.

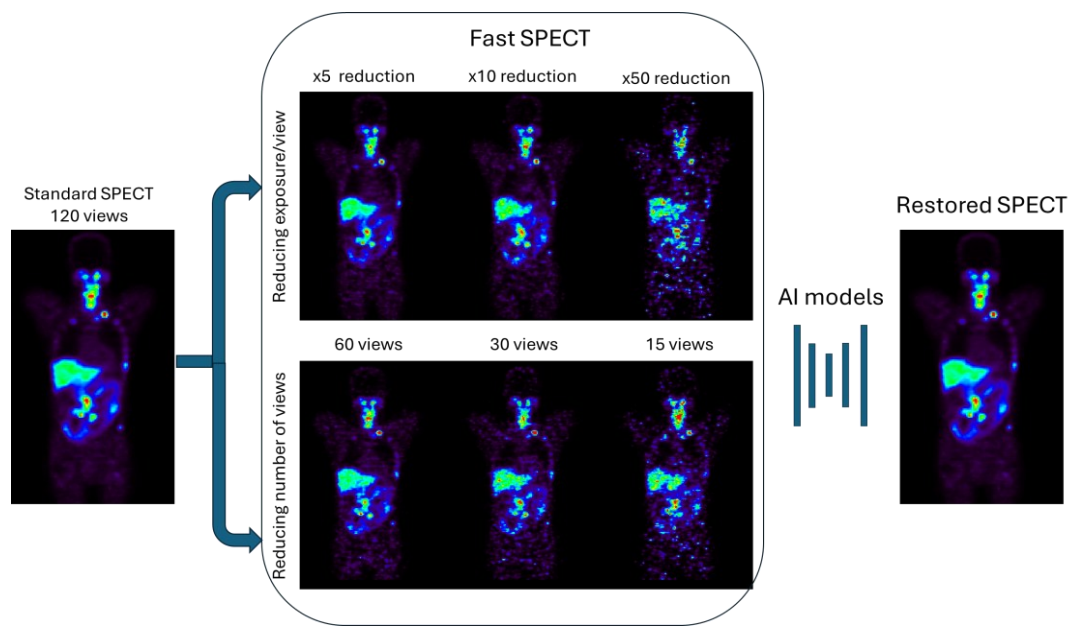

**Fig. 5.** Examples of shortened protocols for fast SPECT acquisitions and impact on image quality, showing degradation for both reduction of exposure per view and reducing number of total views. Also shown is schematic representation of AI models to restore SPECT quality.

*Single time-point dosimetry*

Conventional multi-time-point dosimetry in RPT is requiring serial SPECT/CT or planar acquisitions on multiple days after administration, as recommended by the EANM [73,74]. Existing single-time-point (STP) dosimetry methods rely on mathematical simplification, such as Hänscheid's method for RPT with [177]Lu-DOTATATE/DOTATOC [75], or population-based kinetic parameter estimation, such as Madsen's method for [90]Y-DOTATOC treatment [76]. Other researchers have confirmed the feasibility of these STP methods [77], and reported their high sensitivity to the selected imaging time point [78,79], which restricted their instant usage and application. Wang et al. [80] proposed a data driven approach to reduce sensitivity to time-point selection where they estimated time-integrated activity as a product of measured activity and a nonparametric generalized additive model, significantly reducing the estimation error with the first time point data at day 0 [80]. Vinícius Gomes et al. [81] developed a ML model to predict the effective half-life of the radioligand for different organs and tumors in [177]Lu-PSMA RPT, which then helped calculate the time-integrated activity (TIA) using a mono-exponential decay model [81]. Figure 6 summarizes current STP strategies in RPT.

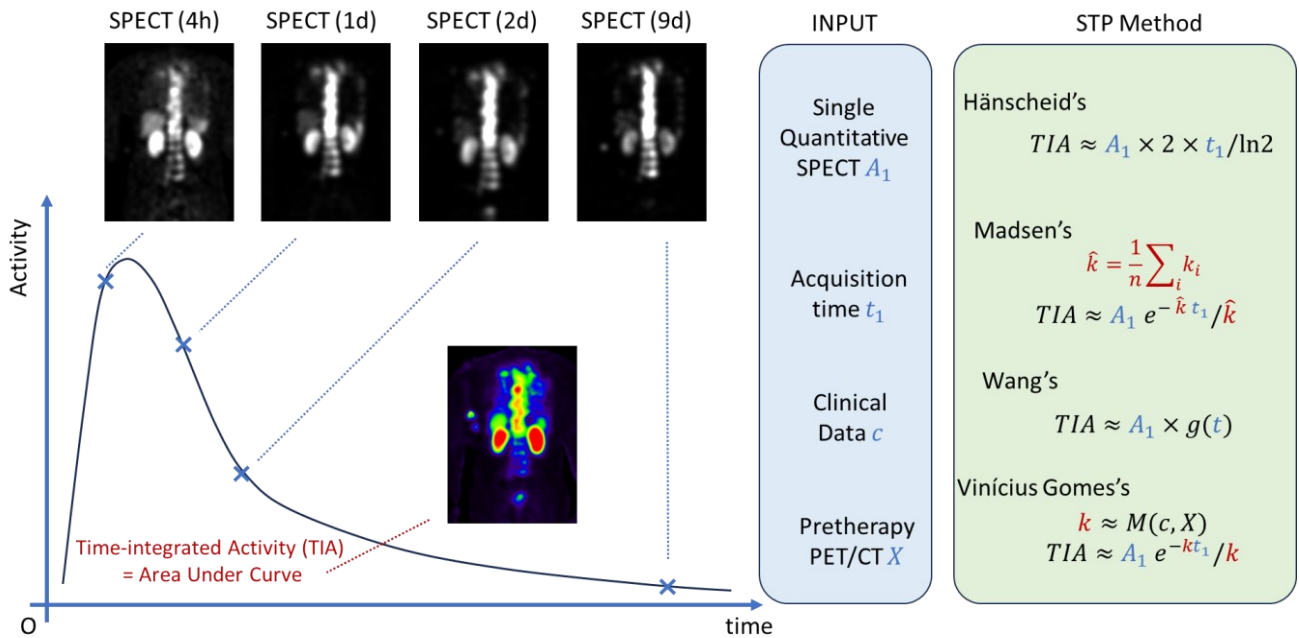

**Figure 6**. An overview of single-time-point dose estimation methods in radiopharmaceutical therapy.

## AI APPLICATIONS TO SIMPLIFY DOSE CALCULATIONS

AI strategies can provide an alternative to Monte Carlo (MC)–based activity-to-dose conversion, alleviating the time and computational burden of MC simulators [51,82] and overcoming the unrealistic assumptions of tabulated S-values or kernel approximations [44]. ML and DL solutions have been proposed to, trained from MC inputs, reproduce similar results [83–87]. For example, Scarinci et al. [87] used AI to reproduce kernel doses. Akhavanallaf et al. [84] included information of density from the CT in a CNN. The authors defined kernels with the activity in the central voxel surrounded by realistic densities to predict voxel-wise dose maps. Combining information about anatomy and activity they reported an improved performance comparable to MC and superior to common solutions based on pre-tabulated methods. Kim et al. [85] used CT images as well as TIA maps as inputs for a UNET model to reproduce the MC dose maps for patients treated with [177]Lu radiopharmaceuticals. Lee et al.[83] used CT and [68]Ga PET images as input to a CNN (UNET) to predict dosimetry in PET imaging. They compared their approach with dedicated MC simulations and S-value kernels showing a comparable performance with MC and a much better accuracy than the kernel approach. Mansouri et al. [86] trained a CNN to generate differences between MC dose maps and multi-kernel dose estimation maps from CT inputs. Those differences are later used to correct the kernel approach solutions and obtain a solution closer to MC estimations.

Other AI solutions not directly related to the conversion of TIA to dose, can have an indirect impact on the quality of that estimation [61]. That is the case of DL methods for accurate image quantification such as attenuation [88–90] and scatter correction [91–93], or resolution limiting factors such as partial volume effects for both SPECT [94,95] and PET [96] or positron range in PET imaging [97,98]. Denoising strategies [99,100] also play a crucial role for accurate quantification dosimetry.

## AI FOR LESION SEGMENTATION

Image segmentation is one of the most common and successful applications of AI [101,102]. Organ segmentation is critical for accurate dosimetry in RPT. DL solutions for organ segmentation have been proposed for CT [103,104] or PET/CT [105,106]. However, lesion segmentation results more challenging due to their variability in shape, size and location. Accurate lesion delineation in RPT enables the extraction of clinically relevant biomarkers and measurement of tumor burden. Manual lesion segmentation remains a significant bottleneck while AI-driven methods offer efficient, reproducible, and accurate segmentation [107,108]. AI-based segmentation techniques provide volumetric biomarkers that are directly associated with tumor burden [109–111], overall survival[112], and patient prognosis [109,113].

Total metabolic tumor volume (TMTV) (Fig.7) is an imaging biomarker that reflects disease burden, aggressiveness, and response to treatment [114–116]. Historically, clinical TMTV quantification has relied on SUV threshold-based techniques[114,117] which suffer from variability and require manual input. AI-based automated TMTV estimation using PET/CT has been validated across several malignancies [118–120], including applications in prostate cancer with PSMA PET/CT [121–123] and in neuroendocrine tumors (NETs) using [68]Ga-DOTATATE [124], [64]Cu-DOTATATE [125], and [18]F-FDG [126,127].

Recent studies have established that AI-derived features offer superior correlation with therapeutic response when compared to standard visually interpreted metrics [121,128,129]. In PSMA PET/CT imaging, segmenting metastatic lesions is complicated by physiological uptake in adjacent organs. While early AI methods were focused on primary lesions [130], recent efforts have shifted toward the automated detection of metastatic sites

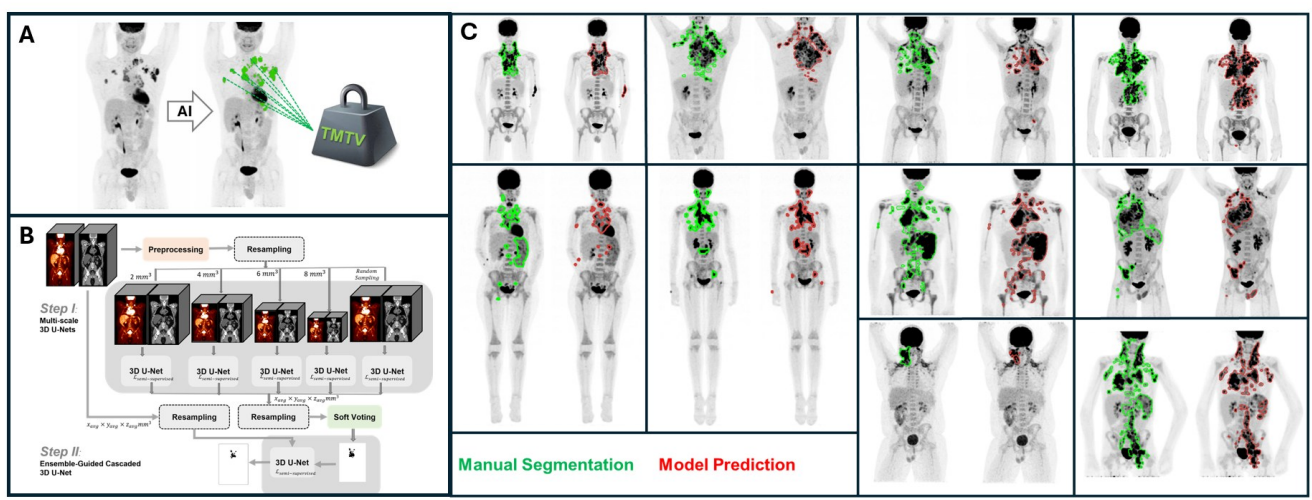

**Fig.7. A.** Schematic definition of Total Metabolic Tumor Volume (TMTV). **B.** TMTV-Net: A fully automated model for TMTV segmentation. **C.** Demonstration of its generalizability from adult training cohorts to pediatric Hodgkin lymphoma patients [115,116]

[131,132] , supporting comprehensive disease characterization essential for personalized RPT. Similarly, in NETs, lesion segmentation in [68]Ga and [64]Cu-DOTATATE PET/CT scans is crucial for capturing tumor burden and heterogeneity, which are directly linked to therapeutic planning. AI models have demonstrated high concordance with expert delineations for liver metastases [133] and overall tumor burden [134], with strong clinical applicability validated in multi-center studies [135].

While DL models have shown strong performance in PET segmentation, they are still under-utilized in SPECT where most methods face challenges in managing lesion appearance/disappearance, deformation, and spatial shifts. This is aggravated due to limited annotated datasets and difficulties generalizing across radiotracers and institutions [136,137]. Some solutions tackle the specific characteristics of SPECT for lesion segmentation. For instance, Fuzzy C-means (FCM) clustering, accounts for voxel intensity and proximity and can be well suited for SPECT's fuzzy boundaries [136].

Despite these obstacles, DL-based segmentation in SPECT/CT has shown potential for facilitating patient-specific dosimetry in RPT [63,137,138]. Transfer learning offers a compelling solution with studies showing that pre-trained models from PET can be adapted for use in SPECT, even across different radiopharmaceuticals with similar bio-distribution patterns [138]. For example, a pilot trial combining [177]Lu-PSMA-617 therapy with stereotactic body radiotherapy (SBRT) [139] demonstrated that lesions segmented from baseline PSMA PET/CT enabled targeted voxel-level dosimetry via SPECT/CT. This opens a scalable path toward generalizable segmentation frameworks in theranostics.

**PRIOR PATIENT INFORMATION TOWARDS AI SIMPLIFIED DOSIMETRY.**

AI methods benefit substantially from incorporating pre-therapy (pre-treatment) patient information. In particular, digital twins—computational patient-specific models that integrate baseline imaging, clinical/laboratory data, and prior dosimetry to simulate treatment response—have emerged as a promising advance in theranostics. In this section, we review how such priors can (i) enhance imaging and quantification (e.g., prior-informed reconstruction/denoising), (ii) improve segmentation and kinetic modeling, and (iii) enable prediction of treatment response and toxicity, thereby supporting patient-specific planning and truly personalized dosimetry.

*AI for Digital Twins*

Theranostic Digital Twins (TDTs) [140] serve as personalized computational models, built from patient-specific data sources such as medical imaging, clinical history, biological markers, and treatment response data (Fig.8). By combining this information, TDTs open the possibility to perform predictive dosimetry (e.g. from pre-

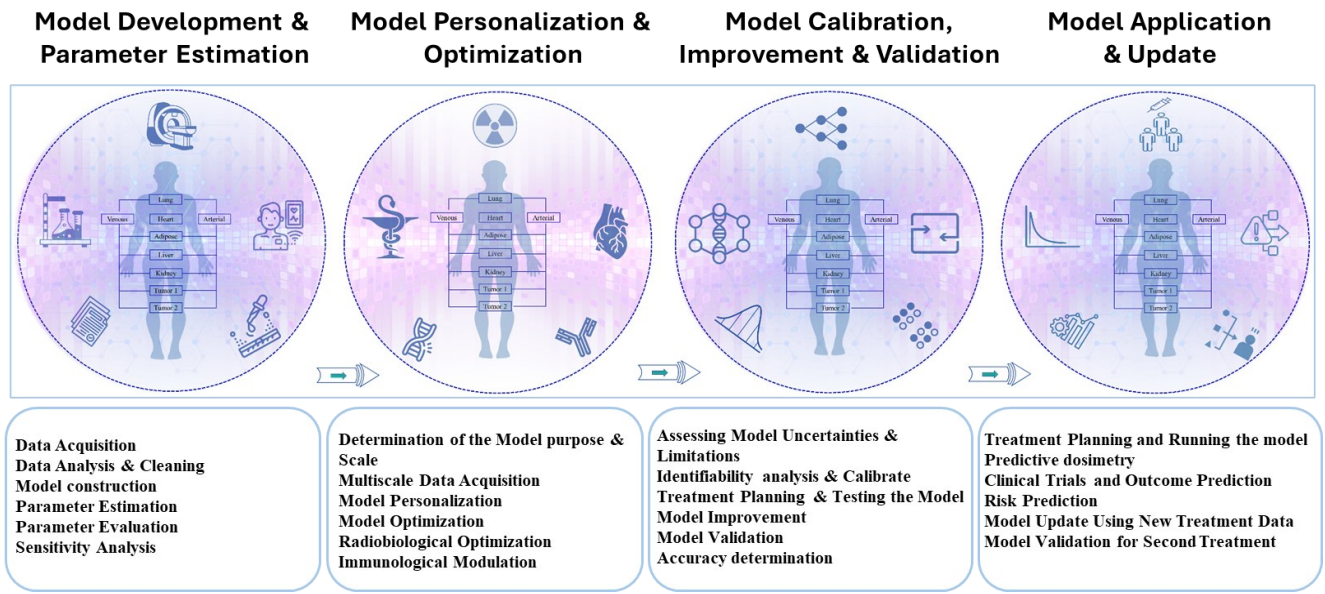

**Model Development & Parameter Estimation**

**Model Personalization & Optimization**

**Model Calibration, Improvement & Validation**

**Model Application & Update**

| | | | |
|---|---|---|---|
| Data Acquisition | Determination of the Model purpose & Scale | Assessing Model Uncertainties & Limitations | Treatment Planning and Running the model |
| Data Analysis & Cleaning | Multiscale Data Acquisition | Identifiability analysis & Calibrate | Predictive dosimetry |
| Model construction | Model Personalization | Treatment Planning & Testing the Model | Clinical Trials and Outcome Prediction |
| Parameter Estimation | Model Optimization | Model Improvement | Risk Prediction |
| Parameter Evaluation | Radiobiological Optimization | Model Validation | Model Update Using New Treatment Data |
| Sensitivity Analysis | Immunological Modulation | Accuracy determination | Model Validation for Second Treatment |

**Fig.8.** The detailed roadmap for the development of theranostic digital twins (TDTs). The main steps are model development, personalization, validation, and application AI can be used for the model development and parameter estimation.[141]

therapy PET scans) and help offer tailored insights for each patient, helping doctors to personalize therapies towards improved treatment outcomes [141]. Advanced modeling techniques like PBPK modeling underpin such development (see companion publications on these topics for this special issue [142–145]). Through computer simulations, these methods help precisely forecast how radiopharmaceuticals distribute within the body, the radiation absorbed by different tissues, expected treatment effectiveness, and potential side effects [146].

To ensure these tools remain practical for routine clinical use, advanced techniques simplify and optimize complex models without losing important physiological insights. Methods such as the Manifold Boundary Approximation Method (MBAM) streamline these detailed models into simpler forms, ensuring ease of use, interpretability, and widespread clinical applicability [147,148]. By combining mechanism-driven physiological modeling with the power and flexibility of AI, digital twins provide scalable, precise, and easy-to-use dosimetry solutions, paving the way for truly personalized radiopharmaceutical therapies.

*From pre-therapy to post-therapy solutions.*

Recent studies have demonstrated that ML can integrate radiomic and dosimetry features from pretherapy $^{68}$Ga-DOTATOC PET/CT and post-therapy $^{177}$Lu-DOTATATE SPECT/CT to predict absorbed doses to OARs, such as liver, spleen, and kidneys, with mean absolute errors below 1.6 Gy [74,149]. Early-cycle imaging data were also found to improve the stability and precision of these predictions, suggesting opportunities for adaptive dosing across therapy cycles [150–152]. Moreover, there is increasing interest in using theranostic pairs (e.g., $^{68}$Ga/$^{177}$Lu) to forecast absorbed doses before initiating treatment. Pre-therapy $^{68}$Ga PET imaging has shown promise for predicting renal dose in $^{177}$Lu PRRT, which is crucial given that renal toxicity is a common limiting factor in PRRT [153]. Notably, renal dose prediction during the first therapy cycle could guide the escalation of administered activity, particularly because the absorbed tumor dose per GBq tends to decrease in later cycles [154]. Approaches that predict therapy dosimetry from pre-therapy PET can be extended to voxel-wise estimation, enabling characterization of intra-organ heterogeneity [55].

In a representative study [155], ML algorithms were applied to predict tumor absorbed dose using pretherapy $^{68}$Ga-DOTATATE PET features and clinical biomarkers. Their approach incorporated multiparametric PET features ($SUV_{mean}$, total lesion burden, liver involvement) and achieved high accuracy ($R^2$ = 0.64, MAE = 0.73 Gy/GBq) in cross-validation. Complementing these findings, a multicenter study evaluated whether kidney absorbed doses from the first $^{177}$Lu-DOTATATE cycle could be predicted from pre-therapy SSTR-PET uptake [156].

Some strategies leverage pre-therapy information to achieve enhanced information from post-therapy 2D planar images. For example, Salimi et al. [157], suggested the use of pre-therapy SPECT/CT images from which they segmented the lungs using DL to predict the lung shunt fraction from planar images after Y90-SIRT therapy. More recently, using pre-therapy patient-specific information in RPT, Lopez-Montes et al. [158] have proposed a method to synthesize 3D SPECT images from only two 2D-planar images leveraging patient-specific datasets of possible uptakes derived from the pre-clinical PET/CT. Finally, dynamic evolution of radiopharmaceuticals can also be leveraged for pre-therapy to post-therapy estimation in dosimetry. In that line of research, Kassar et al. [159] proposed the use of a conditional Generative Adversarial Network (cGAN) with a modified loss term to include PBPK-based dose maps to generate a prediction of the voxel wise dosimetry in $^{177}$Lu-PSMA treatment from pre-therapy $^{68}$Ga-PSMA PET imaging. A schematic representation of models leveraging pre-therapy information for different applications is included in Fig. 9.

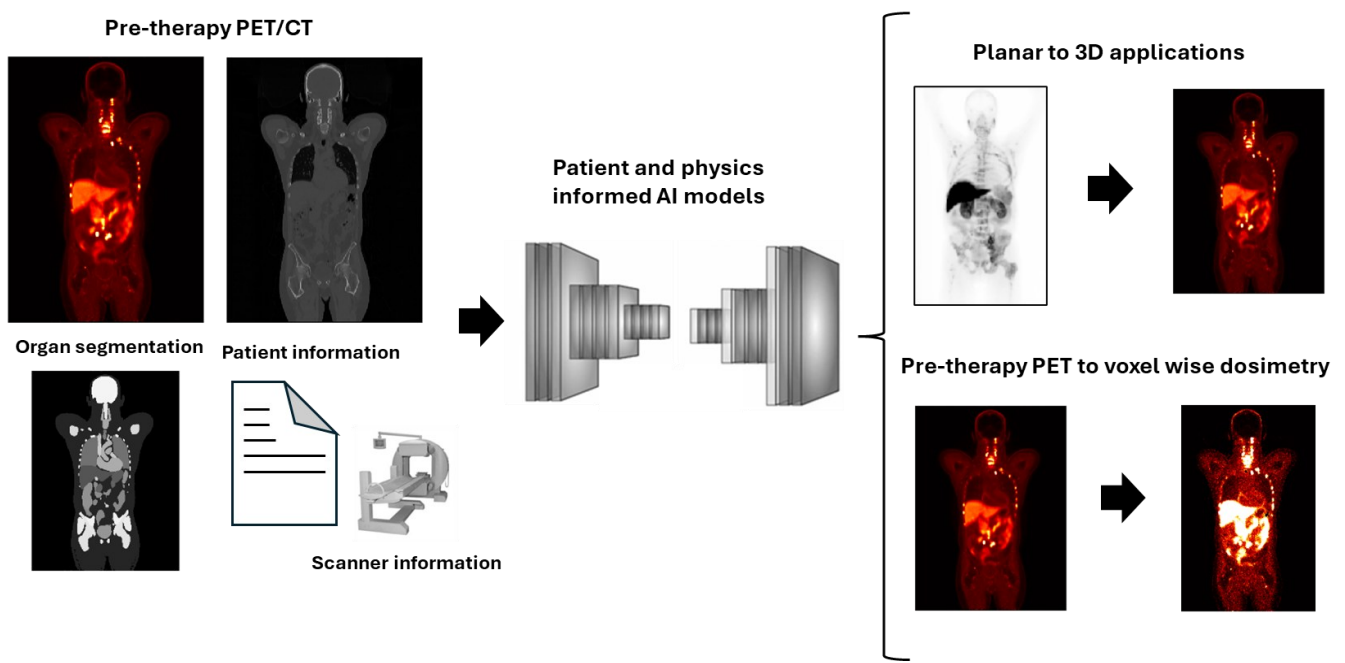

**Fig. 9.** Schematic representations of methodologies to leverage pre-therapy PET and patient information into high fidelity AI towards patient friendly dosimetry.

All these studies collectively illustrate the potential of AI to bridge pretherapy imaging and personalized dosimetry in RPT.

**SUMMARY**

Patient-specific dosimetry is currently a clinical need to evaluate lesion and organs at risk evolution in RPT. Conventional dosimetry protocols are often time and/or computationally intensive which dampers the applicability or real personalized dosimetry. AI strategies can help in the simplification of imaging protocols for dosimetry via reducing the burden of SPECT imaging and/or limiting the number of time-point observations for an accurate estimation of the TIA. DL solutions for TIA to dose conversion present alternatives to costly MC simulations while not relying on generic anthropomorphic models that are agnostic of the patient's anatomy. AI-enabled segmentation strategies support the evolution of personalized, image-guided RPT planning and monitoring. Accurate quantification of radiopharmaceutical uptake and response at the lesion level enable clinicians to assess therapeutic efficacy over time and adapt treatment accordingly. In general, approaches using prior information of the patient to reduce the requirements in both imaging and dosimetry calculations are a major topic that, in our opinion, will be of particular interest within the next years. The merging of pre-therapy information of the patient with the generation of digital twins is a logic evolution of the field that will lead to patient-friendly approaches able to accurately predict outcomes and facilitate administered dose adjustments according to the individual needs of every patient.

While there is still room for enhancement of those techniques to ensure robust, generalizable and verifiable results, AI applications have shown great potential in the field of RPT in general and dosimetry in particular. Collaboration and data sharing between institutions are critical aspects to ensure reproducibility and training with large scale datasets able to generalize population-based models. In this review, we have summarized the main advances in the field of AI, presenting a unique opportunity to tackle personalized, patient friendly dosimetry. These approaches pave the way for a real personalized dosimetry in RPT that will potentially enhance clinical outcomes and improve the quality of life of cancer patients.